# The Shape of Deceit: Behavioral Consistency and Fragility in Money Laundering Patterns

Danny Butvinik, Ofir Yakobi, Michal Einhorn Cohen, Elina Maliarsky

## Abstract

Conventional anti-money laundering (AML) systems predominantly focus on identifying anomalous entities or transactions, flagging them for manual investigation based on statistical deviation or suspicious behavior. This paradigm, however, misconstrues the true nature of money laundering, which is rarely anomalous but often deliberate, repeated, and concealed within consistent behavioral routines. In this paper, we challenge the entity-centric approach and propose a network-theoretic perspective that emphasizes detecting **predefined laundering patterns** across **directed transaction networks**. We introduce the notion of **behavioral consistency** as the core trait of laundering activity, and argue that such patterns are better captured through subgraph structures expressing semantic and functional roles—not solely geometry. Crucially, we explore the concept of **pattern fragility**: the sensitivity of laundering patterns to small attribute changes and, conversely, their semantic robustness even under drastic topological transformations. We claim that laundering detection should not hinge on statistical outliers, but on **preservation of behavioral essence**, and propose a reconceptualization of pattern similarity grounded in this insight. This philosophical and practical shift has implications for how AML systems model, scan, and interpret networks in the fight against financial crime.

## Introduction

In the ever-evolving theatre of financial crime, the act of laundering money remains one of its most elusive performances. What makes it elusive is not its unpredictability, but its **remarkable regularity**. Money laundering, when examined in its operational form, is rarely chaotic. It is structured, procedural, and deliberate—executed with the elegance of choreography rather than the spontaneity of fraud. And yet, most of today's anti-money laundering systems treat it as if it were the latter.

Current practices in AML detection overwhelmingly focus on the identification of **suspicious entities**—individuals or accounts exhibiting behavior that deviates from expected norms. These entities become the focus of attention, and all transactions or connections associated with them are subject to investigation. While this approach has yielded practical results, it carries a fundamental flaw: it presumes that laundering is an *anomaly*, a local deviation. In doing so, it obscures a deeper truth—**that laundering is not defined by deviation but by disguise**, not by anomaly but by **repetition, consistency, and strategic transformation**.

This paper proposes an alternative paradigm, one that shifts the axis of attention from *entities* to *patterns*, from *suspicion* to *structure*, from *statistical deviation* to **semantic persistence**. At the heart of this view lies the network representation of financial systems—directed graphs where nodes represent entities and edges transactions. Within these graphs, laundering reveals itself not as noise, but as **repeated motifs**: subnetworks that encode obfuscation, layering, integration, and circular movement. Detecting these motifs requires not anomaly detection but pattern recognition.

Yet pattern recognition in graphs is not straightforward. It is not enough to find geometric matches of subgraphs. A laundering structure can remain **topologically identical** while losing all criminal relevance due to a minor change in an attribute—say, a corporate node replaced by an individual account. Conversely, two patterns might appear **geometrically dissimilar**, but if they functionally achieve the same laundering objective, they should be considered equivalent. This disjunction reveals a central tension in pattern-based AML: the fragile boundary between sameness and difference.

We call this the **fragility of laundering patterns**—the idea that laundering behavior is simultaneously easy to distort and difficult to eliminate. Fragile patterns can be broken with minimal changes; robust ones can morph substantially while preserving their essence. This conceptual axis—fragility versus robustness—becomes critical in understanding how to define, compare, and detect laundering behaviors in practice.

This paper explores that axis. It asks: **What makes two laundering patterns the same? What makes them different? How do we define "matching" in a domain where behavior hides in the interplay of form and function?** To answer these, we propose a shift from geometric subgraph matching to *semantic* and *behavioral matching*, wherein the focus lies not on whether two graphs look alike, but whether they behave alike—whether they express the same laundering intent through different transactional choreography.

This rethinking, we argue, is not merely theoretical. It has immediate consequences for how AML systems are designed, how investigations are triaged, and how intelligence is automated. If we are to detect laundering not by catching suspects, but by recognizing behaviors, we must build systems attuned to the **shape of deceit**—to the subtle, structured, yet fragile forms in which financial crime conceals itself.

## Directed Networks and the Grammar of Money Laundering

To understand money laundering as a patterned behavior, rather than a localized anomaly, we must begin with the language in which such behaviors are expressed. This language is **not statistical** in nature, but **structural and directional**. It speaks through networks, and more precisely, **directed networks**—graphs in which each edge has a direction, indicating the flow from one node to another. In the financial domain, these graphs are not abstract constructs; they are exact representations of reality. Each node corresponds to an **entity**—a person, an account, a company—while each directed edge denotes a **transaction**, complete with a source, a destination, and attributes like value, time, and type.

In this setting, the direction of edges is not a convenience—it is an ontological necessity. Money laundering is a flow-driven process. The essence of the act is to take illicit value from an origin, pass it through intermediaries and layers, and reintroduce it into the financial system under the guise of legitimacy. This journey cannot be captured in an undirected world. It requires direction to encode causality, agency, and transformation. The edge from A to B is not equivalent to the edge from B to A. One speaks of *sending*, the other of *receiving*, and in AML, this distinction is fundamental.

But the richness of directed networks lies not just in their capacity to record individual movements of money. Their true power emerges when we treat them as grammatical systems. Just as natural language strings together words into coherent statements, laundering patterns string together transactions into financial narratives. Some are simple clauses—a transfer followed by a withdrawal. Others are more complex—multi-entity paths designed to confuse, obscure, and fragment the traceability of value. These are the sentences of laundering, and the network is the grammar that governs them.

In this grammar, laundering operations form **subgraphs**—coherent, bounded regions of the network that collectively express a laundering behavior. These subgraphs are not defined solely by topology, but by roles. One node serves as the entry point for illicit funds, another as a layering intermediary, a third as the integration endpoint. The edges are verbs: moving, splitting, combining, cycling. And just as in language, the same sentence can be rephrased with different words or structures but preserve the same meaning, laundering patterns can undergo transformation while maintaining their functional role.

Consider two subgraphs. One is a classical funnel: multiple accounts send money to a central node, which then forwards it to a final recipient. The second is a daisy-chain: the same value hops sequentially through a line of accounts before arriving at the same destination. Topologically, they differ—one is fan-in, the other is linear. But both perform the same grammatical act: consolidation and redirection of funds. From a laundering perspective, they are semantically equivalent.

This insight challenges conventional pattern-matching approaches, which focus on *shape*. It suggests instead that we must understand *function*—what the pattern *does* within the directed graph. The directionality of the network becomes the backbone of this understanding, allowing us to distinguish flows of value, reconstruct intentions, and detect **behavioral invariants**.

Thus, directed networks are not mere data structures; they are semantic engines. They encode the grammar of money laundering—the syntax, the verbs, the clauses through which illicit financial behavior expresses itself. And it is within this directed grammar that laundering patterns reside, mutate, hide, and—when properly understood—reveal their presence.

# From Geometry to Semantics: Rethinking Pattern Matching

In the standard graph-theoretic paradigm, pattern matching is primarily concerned with structure and topology. A subgraph is matched if its geometry—its shape, size, and edge configuration—aligns with a known template. Algorithms for subgraph isomorphism or approximate matching are designed to find such geometric congruencies. This approach, while computationally sound and practically useful in many domains, fails to capture the essence of functional behavior, especially in the context of money laundering.

Money laundering patterns are not merely architectural. They are intentional, and their expression in a network is driven by purpose, not symmetry. A laundering scheme may reuse certain topological motifs—a star, a chain, a ring—but the structure itself is secondary. What matters is how entities interact, roles are played, and flows are manipulated to achieve obfuscation and reintegration of illicit value. The structure is the *syntax*; the laundering behavior is the *semantics*.

To illustrate, consider two subgraphs: both contain three nodes and two directed edges. In the first, a business account (A) sends funds to an individual account (B), which immediately forwards them to another business account (C). In the second, the same structure exists—A → B → C—but the node types differ: now, A and C are both personal accounts, and B is a corporate shell. From a purely geometric perspective, the graphs are indistinguishable. But semantically, they are profoundly different. The first may reflect payroll or vendor payments; the second could encode a classic layering tactic. **Geometry is identical; meaning is not.**

Conversely, two graphs can differ substantially in topology—say, one forms a star and the other a chain—but still achieve the same laundering goal: taking value from multiple sources, obscuring its trail through intermediaries, and reintroducing it under a unified output. Here, the **semantics are preserved despite structural divergence**. This is akin to two sentences that say the same thing using entirely different grammatical constructions. What binds them is not shape, but **intent and effect**.

This disconnects between **geometric similarity** and **behavioral equivalence** lies at the heart of the challenge. Traditional pattern matching in graphs assumes that similar geometry implies similar meaning. But in laundering detection, this assumption is not only flawed—it is dangerous. It leads to **false negatives** when patterns that look different act the same, and **false positives** when patterns that look the same behave differently.

To overcome this, we must reconceive what it means to "match" a pattern. Instead of asking *does this subgraph look like the template?* we must ask *does this subgraph behave like the template?* This shift demands that we move beyond structural isomorphism and toward **semantic congruence**: a deeper alignment of roles, flows, and intentions encoded in the graph. It means treating attributes not as mere decorations on nodes and edges, but as *critical features* that shape the meaning of a pattern.

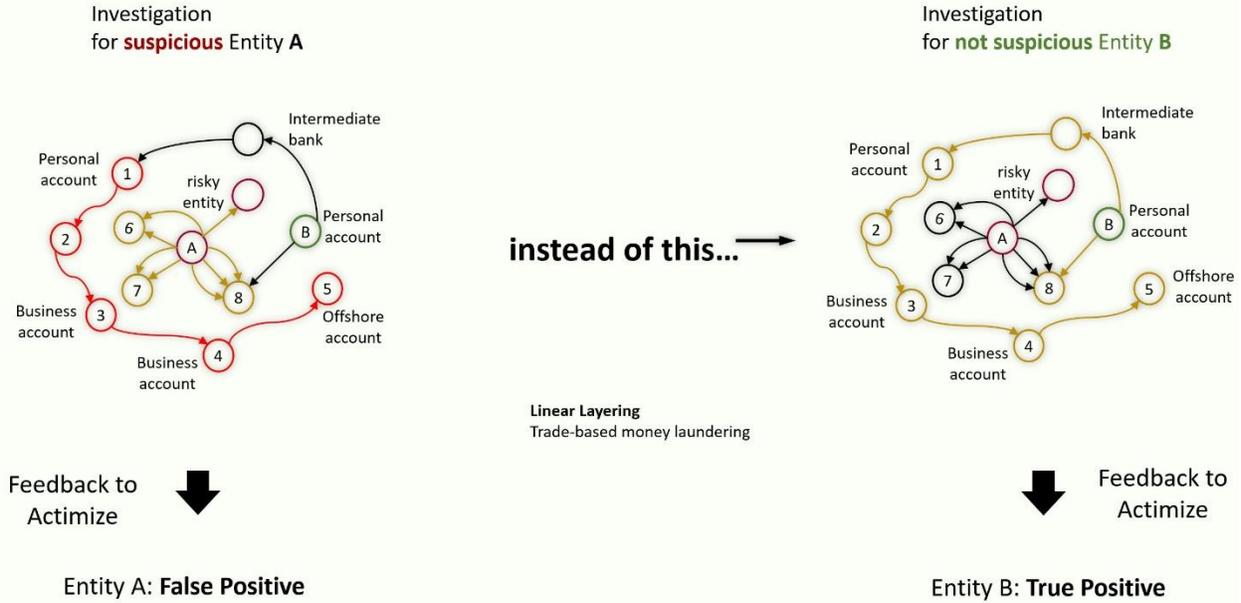

Figure 1: Case Management Investigation — Non-indicativeness of Anomalies and the Primacy of Behavioral Pattern Detection

This figure illustrates a critical failure of current anomaly-based AML systems: the misalignment between statistical suspicion and actual illicit behavior. It contrasts two investigative paths: the left-hand side triggered by a suspicious entity (Entity A), and the right-hand side based on a consistent laundering pattern that involves a non-suspicious entity (Entity B).

Left (Entity A - False Positive)
Entity A has been flagged as suspicious due to its anomalous features—possibly due to transaction frequency, value deviation, or external risk classification. This leads to a full investigation centered around it. However, the surrounding subgraph, while irregular in structure, does not manifest any coherent laundering behavior. The entity serves as a transactional hub, but without clear layering, integration, or obfuscation logic. The result is a false positive, a waste of investigative resources, and a missed opportunity to uncover actual illicit operations.

Right (Entity B - True Positive)
In contrast, Entity B is not anomalous in itself—its attributes and behavior fall within expected statistical bounds. However, it participates in a well-formed laundering pattern, evident in the linear and semantically consistent layering path surrounding it. This includes the movement of value through business and offshore accounts in a directed and plausible laundering sequence. Despite B's innocuous local appearance, it plays a role in a global illicit structure. Recognizing this yields a true positive, but only if one looks beyond isolated anomalies and considers entire behavioral subgraphs.

Conclusion

The figure sharply reinforces your thesis: entities are misleading as the unit of suspicion. What matters is not local anomaly, but participation in behavioral archetypes. This visual demonstrates the core danger of entity-centric AML—its inability to see the forest of laundering patterns for the trees of suspicious accounts. The feedback loop into the system further solidifies this problem: false positives are dismissed, while true laundering patterns go undetected due to their non-anomalous nodes.

This reframing also has implications for how we define pattern libraries. A laundering pattern is not a static diagram but a **behavioral archetype**—a function over a space of possible realizations. It defines not a single geometry, but a **family of semantically equivalent subgraphs**, each varying in size, configuration, or attributes, yet all preserving the same illicit logic. Matching, then, becomes the task of identifying whether a subgraph belongs to this family—not whether it looks identical to a specific exemplar.

In this way, we move from the **geometry of graphs** to the **semantics of financial behavior**. And in doing so, we open the door to a more nuanced, resilient, and meaningful approach to money laundering detection—one that respects the fragility of surface forms and the durability of behavioral essence.

## The Fragility of Illicit Pattern Identity

Money laundering patterns are deceptive not only in intent, but in their **ontological instability**—their ability to retain or lose meaning through minimal alterations. This introduces a rarely addressed but fundamental problem in detection: **how fragile is a laundering pattern's identity?** In other words, when does a laundering pattern stop being one? What amount of change—structural, directional, or attributive—is sufficient to preserve, distort, or entirely dissolve its illicit essence?

We introduce the notion of **pattern fragility** to capture this phenomenon: the sensitivity of a money laundering pattern to perturbations in its expression. Fragility, in this context, is not about noise or randomness. It is about the **semantic dependency** of a behavior on specific features—particularly the attributes of nodes and edges, the roles they assume, and the logic of their connectivity. A fragile pattern is one where a subtle change (a node type switch, a small shift in transaction value, or a single directional flip) can completely transform its interpretative meaning—from illicit to benign. A robust pattern, by contrast, can undergo substantial transformations—growing in size, changing topology—while still enacting the same laundering operation.

Consider a simple archetype: a corporate account sends funds to a series of personal accounts, which in turn converge into a second company. This fan-out followed by fan-in may signify a layering process designed to obfuscate origin. Now, suppose one node in this configuration changes its type from "personal" to "business," or the transaction metadata shifts from "cash" to "internal transfer." The graph structure remains, the directionality unchanged—but the meaning collapses. What was once a laundering pattern now resembling an intercompany transfer scheme. This is **semantic collapse through minimal perturbation**: high fragility.

Now contrast with another pattern—say, a series of wire transfers executed between multiple offshore accounts, across jurisdictions, through seemingly unrelated intermediaries. The topology here can be changed in many ways—entities reordered, paths split or lengthened, parallel flows inserted—and yet the functional identity of the pattern remains intact. The purpose—to distance origin from endpoint through high-velocity jurisdictional fragmentation—is preserved. This is **semantic resilience through structural flexibility**: low fragility, high robustness.

The concept of fragility forces us to rethink what we often take for granted in pattern-based detection: that patterns are well-defined, that they can be matched via templates, that slight variations are irrelevant. But in laundering, **variations are often the entire story**. Criminals exploit the detection systems' reliance on rigid pattern matching by introducing minimal alterations—small enough to evade templates, yet subtle enough to preserve laundering function. They **engineer fragility**, turning it into a tactical advantage.

This fragility is compounded by the **multiplicity of equivalence**: multiple structures can enact the same function, but the same structure can instantiate multiple, incompatible functions. This is a non-bijective relationship between form and behavior—**many-to-one and one-to-many mappings**—that makes any geometric approach to detection inherently brittle.

Detecting laundering, then, is not about spotting repeated shapes. It is about capturing **semantic invariants** under a spectrum of distortions. A pattern should be treated as an equivalence class—not a fixed subgraph, but a flexible semantic shell, capable of manifesting in diverse forms yet retaining its core laundering logic. The challenge is to distinguish meaning-preserving perturbations from meaning-erasing ones, and to quantify the **tolerance** of a pattern's identity to those changes.

This introduces a rich and underexplored axis of analysis: the **distance between a known laundering pattern and a candidate subgraph**, not in terms of edge edits or node substitutions, but in terms of *functional deviation*. How much semantic drift is allowed before we no longer consider a behavior to be laundering? What makes a pattern fragile, and can that fragility be formalized, measured, and used as a filter or risk amplifier?

These questions are more than technical. They cut to the epistemic foundation of AML systems: how do we define what laundering is, and how do we know when we've seen it? Pattern fragility challenges our assumptions, exposes our blind spots, and demands that we move beyond rigid templates toward fluid, behaviorally aware, semantically grounded representations of financial crime.

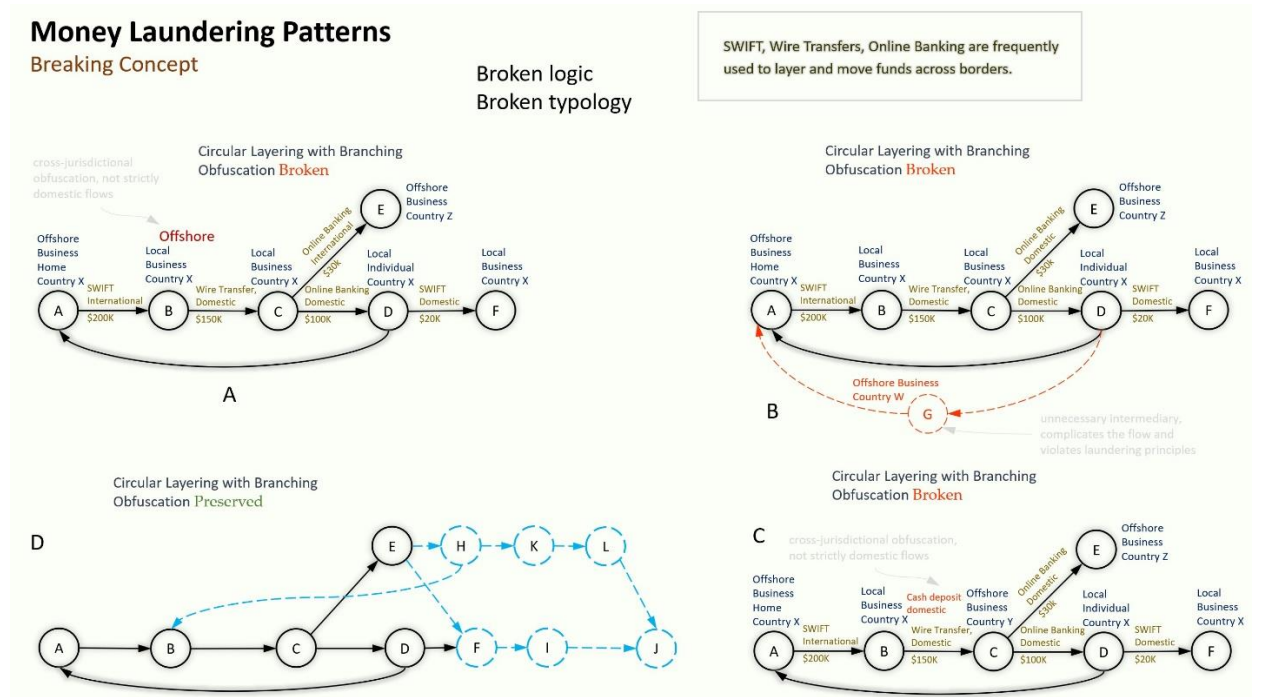

Figure 2: Fragility of Laundering Patterns — Preserving or Breaking the Semantic Essence Across Structural Variants

This figure addresses the core concept of pattern fragility, by showing how laundering structures can either break under minor changes or persist through substantial transformations.

Top Row: Patterns A, B, and C — Broken Concept
Patterns A, B, and C are geometrically consistent but semantically compromised. All three attempt to implement circular layering with branching, but the laundering concept is broken in different ways:

- Pattern A includes entities and flows that violate laundering logic—for example, unnecessary intermediaries, jurisdictional inconsistencies, or entity types that invalidate the obfuscation function.

- Pattern B introduces an offshore business node (G) that overcomplicates the circular logic and exposes the pattern as contrived. The pattern is no longer elegant or plausible from a laundering execution standpoint.

- Pattern C breaks the logic through a cash deposit, which disrupts the layering narrative and introduces traceable, high-risk behavior uncharacteristic of laundering mechanisms aiming for obfuscation.

These examples demonstrate semantic fragility: despite structural resemblance, the behavioral integrity collapses with small changes. This fragility is not visible in the geometry—it emerges only through behavioral interpretation.

Bottom Row: Pattern D — Preserved Concept Despite Structural Expansion

Pattern D offers a stark contrast. Its structure is more complex and differs significantly from the canonical circular form. However, it preserves the laundering logic. The extended path (nodes H–L) branches the obfuscation process, increases distance, and diffuses traceability—but crucially, does not violate the semantic core. The roles remain functionally aligned, jurisdictional separation is maintained, and the layering remains plausible.

This illustrates semantic robustness: the pattern sustains its illicit logic despite architectural divergence. From a topological viewpoint, D is less similar to the canonical template than A, B, or C—yet it is the only valid laundering archetype among them.

Conclusion

Figure 2 encapsulates the core of your fragility thesis: illicit pattern identity is not guaranteed by structural similarity, nor is it lost by structural variation. The distinction between matching in form and matching in function becomes critical. These examples force the AML system to evolve from geometry-based reasoning to semantically grounded detection—a perspective that tolerates visual drift but insists on behavioral fidelity.

## The Behavioral Essence of Laundering Patterns

If the structure of a transaction network is its grammar, then laundering is not the syntax—it is the story. A story of transformation, not anomaly. It begins with tainted value and ends with its reappearance as clean. In between lie transactions, entities, delays, and misdirection, but more fundamentally, lies purpose. To detect laundering, we must grasp not just how the network is wired, but why it is wired that way. And that "why" resides in **behavioral essence**—the function a pattern enacts, regardless of its structural dress.

The essence of a laundering pattern cannot be captured solely through topology, nor through isolated attributes. It is expressed through **a choreography of interactions**: entities play roles, transactions perform acts, and the system as a whole achieves a transformation from illegitimate to legitimate financial appearance. Some entities split funds; others layer them; still others reintegrate them through innocuous endpoints. These behaviors are not always anomalous—indeed, that is their genius. They are often locally plausible and globally orchestrated.

Thus, the heart of a laundering pattern is **not where the red flag appears**, but **where the transformation occurs**. This shifts the locus of detection from points (suspicious accounts) to processes—how value morphs and migrates over time and across connections. This is why a structurally different pattern may still instantiate the same behavior: if the path, timing, and role logic are preserved, the laundering mechanism persists.

Take two distinct subgraphs: one uses five intermediary accounts to layer and obscure; another uses just two, but across higher-risk jurisdictions with rapid-fire transfers. Structurally, they differ; in behavior, they are twins. Both perform semantic laundering—altering the appearance, traceability, and perceived legitimacy of value. This behavior is the invariant—the essence that must be captured.

Detecting such patterns thus requires models that are **not sensitive to superficial form**, but attuned to functional logic. One cannot rely on simple heuristics like "three hops" or "fan-in equals suspicious." Instead, one must define and detect **behavioral fingerprints**—combinations of flow, role, timing, and metadata that together enact laundering. These are not static features, but dynamic, interacting dimensions that constitute meaning.

From this perspective, each laundering pattern is an **operational archetype**, a strategy rather than a shape. A pattern is not defined by its number of nodes or edges, but by the **narrative it enacts over the graph**. It is a latent function—observable not through snapshots, but through movement. Thus, the true challenge of AML is not to define laundering in terms of subgraph isomorphism, but to encode and detect **recurrent behavioral invariants**—semantic signatures of illicit transformation that persist beneath the surface variation.

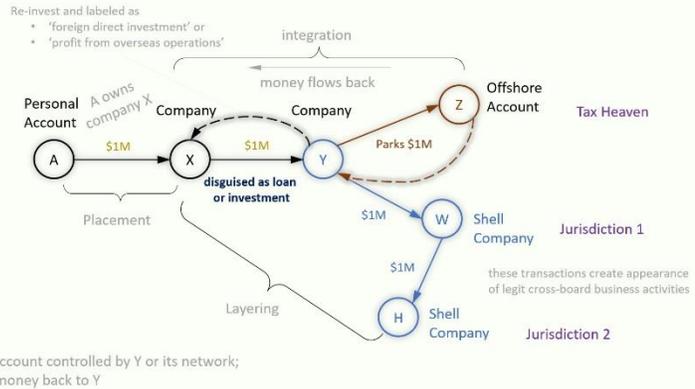
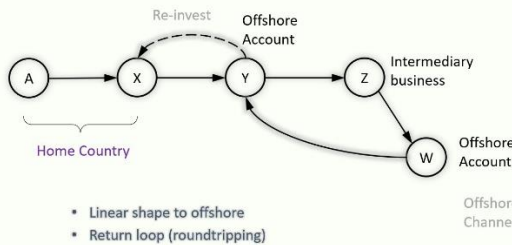

Figure 3: Round-Tripping Layering — Circular Behavioral Logic Disguised as Cross-Border Investment

This figure captures a particularly subtle and semantically robust laundering **archetype**: round-tripping, where illicit funds are exported and subsequently reintroduced into the originating economy disguised as legitimate income—often under the cover of foreign direct investment or business profit. This typology is frequently associated with tax evasion, capital flight, and *regulatory arbitrage*, and it leverages cross-jurisdictional opacity to transform the illicit nature of funds into an apparent economic contribution.

Left Sub-Figure: Structural Overview of Round-Tripping

This network representation shows the basic laundering choreography:
- A **personal account (A)** initiates the flow by transferring funds to **Company X**, which is domestically registered but often controlled by the same individual.

- Funds then move linearly through a chain of companies—**Y** and **Z**—ultimately reaching an **offshore account** or jurisdiction (e.g., Cayman Islands, Panama).

- The offshore account (Z or W) is often legally or beneficially controlled by the same originating party or its proxies.

- Finally, the money **returns to the original source**, or its economic equivalent, under a legitimate label (e.g., loan, equity, or foreign profit).

This shape is linear at first, but contains a hidden behavioral loop—a return path masked beneath a web of intermediaries and international boundaries. Structurally, the graph may not look like a cycle; semantically, it is circular in intent and effect. This is a prime example where geometry misleads, and only behavioral interpretation uncovers laundering logic.

Right Sub-Figure: Annotated Semantics and Jurisdictional Layering

The right-hand figure delves deeper into the *semantic roles and jurisdictional manipulation* that give this pattern its laundering potency

- **A → X → Y** represents the classic placement phase: an individual (A) channels personal funds into a domestic company (X) they own, which then transfers them to another domestic or semi-domestic entity (Y), labeled as a business partner or subsidiary.

- From **Y → Z**, the money enters a tax haven (e.g., through an offshore account). Importantly, **Z is not a final sink**—it is a parking entity, not a laundering endpoint.

- The money is routed back to the system via **W and H**, which are *shell companies* created in other jurisdictions (Jurisdiction 1 and 2). These shells do not perform real economic functions; they exist solely to simulate legitimate financial transactions—loans, returns, reinvestments, or capital flows.

- The final integration occurs when the funds return to entity Y, now recharacterized as *foreign income*, *business return*, or *investment inflow*—taxed differently or not at all, and entirely disassociated from the original illicit source (A).

This behavior illustrates the tri-phasic laundering structure:
1. Placement (A to X/Y),
2. Layering (Y to Z via shell intermediaries),
3. Integration (Z or W back to Y or X, often disguised as a reinvestment).

The jurisdictional complexity is a critical laundering enabler here:

- Each movement crosses borders, often into *offshore financial centers* with minimal regulatory disclosure.

- Shell companies in Jurisdictions 1 and 2 act as behavioral buffers, adding semantic distance between the initial source and the integrated result.

- Importantly, **none of these flows are inherently anomalous**. They mimic legal behaviors—international loans, foreign investments, intra-group transfers—which makes them invisible to systems focused solely on entity anomalies or transaction thresholds.

Conceptual Implications

This figure is essential for several reasons:
- It shows how non-circular structures can enact circular behaviors—the "round-trip" is semantic, not topological.

- It demonstrates the **role of disguise**: laundering here is not about hiding transactions, but about relabeling them through functionally neutral transformations.

- It affirms your core thesis: money laundering is behavioral transformation, not statistical deviation.

- It introduces the concept of jurisdictional semantics: the meaning of a transaction depends not just on who and how, but *where*. Crossing regulatory regimes is itself a laundering function.

Conclusion

Figure 3 adds depth to your exploration of laundering as **semantic choreography across directed networks**. It vividly shows how **intent**—not appearance—defines the laundering act. By walking through the stages of round-tripping, this figure shows that detection must not rely on structural templates, but on **recognizing functionally equivalent behavioral loops**, even when hidden behind linear facades and jurisdictional camouflage.

## Conceptual Distance and Pattern Equivalence

If we accept that money laundering patterns are behavioral in essence, then we must address a central, unresolved question: **when are two patterns the same?** Not syntactically or geometrically, but conceptually. What defines *equivalence* among laundering behaviors, and how can we measure *distance* between candidate subgraphs and known illicit templates?

Classical graph theory offers precise metrics for similarity—edit distance, graph kernels, spectral embeddings. But these capture form, not function. They tell us how many edges must be removed or nodes relabeled to align two graphs—but they say nothing about whether the same laundering strategy is at work. In AML, this leads to both false matches (structurally similar but semantically benign) and missed matches (structurally different but behaviorally identical).

To move beyond this, we must develop a notion of **conceptual distance**—a measure of behavioral deviation. This distance is multi-dimensional, incorporating:

- **Entity role alignment**: Are the same functional roles (originator, splitter, layer, integrator) enacted, even if by different node types?

- **Flow logic**: Does the subgraph preserve the same temporal and directional transformation of value?

- **Transaction semantics**: Are the transaction types, amounts, frequencies, and channels functionally comparable?

- **Jurisdictional or regulatory overlays**: Does the risk landscape resemble the reference pattern?

These dimensions form a **semantic manifold**, within which each subgraph can be embedded. The closer a candidate is to a known laundering archetype in this space, the higher its behavioral similarity—regardless of surface form. In this view, a laundering pattern is not a static object, but a **region in conceptual space**. Matching becomes a matter of **semantic proximity**, not syntactic replication.

This also clarifies the idea of **equivalence classes**: multiple topologies, when projected into this behavioral space, collapse into a single functional identity. Just as different code snippets can implement the same algorithm, different subgraphs can encode the same laundering logic. This allows us to define robust pattern families, not brittle templates.

At the same time, we gain the ability to measure deviation—to quantify how far a behavior strays from a known laundering archetype. This can support fuzzy matching, confidence scoring, and gradient risk assessment. It also opens the door to **novelty detection**: when a subgraph lies near multiple archetypes but matches none precisely, it may represent a variant laundering strategy, previously unseen but semantically close.

Ultimately, conceptual distance reorients detection from *matching diagrams* to *recognizing behaviors*. It acknowledges the polymorphic nature of laundering, respects its semantic fluidity, and equips us to detect its invariants across shape, size, and surface deception. It is this distance— measurable yet meaning-driven—that should form the foundation of modern AML pattern detection.

Interestingly, this idea of behavioral equivalence resonates with how computational linguistics treats structural variance through the lens of shared generative rules. In formal grammar theory, for instance, multiple syntactic constructions can be derived from the same underlying rule in a context-free grammar. Similarly, in the laundering domain, multiple subnetworks — each differing in shape, length, or composition — may in fact be concrete manifestations of the same abstract laundering routine, defined by its procedural roles: placement, layering, and integration. Just as different linguistic spans converge to the same derivational schema, laundering structures that differ geometrically may instantiate identical behavioral intent. This analogy helps reinforce the distinction between topological similarity and semantic identity that lies at the heart of conceptual distance.

## Towards a New Philosophy of Detection

The prevailing paradigm in anti-money laundering systems remains rooted in the logic of anomaly and suspicion. Suspicious activity reports (SARs), risk scores, and entity monitoring frameworks all converge on a single operational assumption: that laundering reveals itself through *deviation*—through entities or transactions that stand out against normative baselines. This paradigm, while useful in early generations of compliance technology, is increasingly inadequate in facing the evolving sophistication of financial crime.

Laundering today does not stand out; it blends in. It adopts the garments of legitimacy and weaves itself seamlessly into the transactional fabric of modern finance. Its innovation lies not in its ability to violate norms but to mimic them—to operate beneath the statistical radar by distributing risk, flattening patterns, and exploiting the blind spots of detection systems that look for noise instead of narrative.

A core modeling difficulty in this context arises from the fundamental lack of canonical ordering in graph-structured data. Unlike text or images, where spatial or sequential order guides the learning process, graphs exhibit permutation invariance — they are agnostic to node traversal order. However, money laundering is inherently procedural, exhibiting an implicit behavioral flow from placement to layering to integration. This disconnect poses a challenge to existing GNN architectures, which often fail to encode such directed behavioral semantics. One possible resolution lies in semantically structuring the graph traversal process, injecting domain-relevant order into the learning pipeline — for example, defining entry points (placement nodes) and propagating context directionally through layering and integration. This reflects a deeper shift: from learning **on** graph structures to learning **through** the behaviors they encode.

What this paper proposes is not a new technique, but a new philosophy—a foundational rethinking of how laundering should be understood, modeled, and ultimately detected. At the core of this philosophy is a conceptual inversion: **laundering is not anomaly, it is repetition; not outlier, but strategy; not statistical surprise, but behavioral consistency cloaked in variation.** Detection, therefore, must transition from the search for deviant data points to the recognition of functional patterns—recurrent motifs that preserve the *essence* of illicit transformation across surface diversity.

This philosophy places **patterns—not entities—at the center of AML efforts**. It suggests that instead of flagging a suspicious account and launching an investigation into all its adjacent transactions, one should scan the entire network for expressions of laundering logic, regardless of which nodes enact it. The unit of detection is not a person, not a transaction, but a **semantic** **structure and network topology**: a set of roles, flows, and transformations that together enact the laundering act.

It also challenges the false dichotomy between structure and semantics. In this new paradigm, a pattern is neither defined purely by its topology nor its attributes alone, but by its ability to functionally reproduce the laundering process—to carry illicit value across layers of abstraction and deliver it safely into the legitimate economy. Detection systems must therefore shift toward behavioral modeling, embedding semantic understanding directly into pattern libraries, detection algorithms, and risk evaluation frameworks.

This move also necessitates the abandonment of rigid templates in favor of flexible, probabilistic**,** and context-aware models of pattern equivalence. Instead of binary matching, we must embrace **gradual similarity**, conceptual distances, and fuzzy boundaries that reflect the real-world fluidity of criminal innovation. Laundering is not a static adversary; it is adaptive, expressive, and resilient. Our detection philosophy must be the same.

Finally, this new philosophy reclaims an important epistemic stance: **not everything suspicious is laundering, and not all laundering is suspicious.** What unites laundering behaviors is not that they look strange, but that they achieve a transformation under the cover of legitimacy. It is this behavioral transformation—camouflaged, distributed, patterned—that we must learn to detect. In embracing this shift, we are not merely improving AML performance. We are aligning our models with the *ontological truth* of laundering as an adversarial, semantically rich, structurally fluid behavior. And in doing so, we build systems that no longer chase shadows of suspicion, but pursue the *shapes of deceit*—patterns whose meaning is deeper than their geometry and whose detection begins only when we truly understand their intent.